\documentclass[journal, letterpaper, twocolumn]{IEEEtran}
\usepackage{citesort}
\usepackage{bm}
\usepackage{color}
\usepackage[latin9]{inputenc}
\usepackage{amsthm}
\usepackage{amsmath}
\usepackage{amsthm}
\usepackage{amssymb}
\usepackage{graphicx}
\usepackage{lipsum}
\usepackage{algorithmic}
\usepackage{algorithm}
\usepackage{mathrsfs}
\usepackage[table]{xcolor}

\theoremstyle{definition}

\theoremstyle{remark}

\newcommand{\comment}[1]{ }

\newcommand{\PU}{\ensuremath {\mathit{PU}}}
\newcommand{\n}{\ensuremath {\mathit{n}}}
\newcommand{\SU}{\ensuremath {\mathit{SU}}}
\newcommand{\nsamples}{\ensuremath {\mathit{L}}}

\newcommand{\dfti}{ {\hat{\boldmath{F}}} }
\newcommand{\measure}{ {{\boldmath{\Phi}}} }
\newcommand{\measuree}{ {{\boldmath{\phi}}} }
\newcommand{\sense}{ {{\boldmath{\Psi}}} }
\newcommand{\Winv}{ {{\boldmath{W}^{-1}}} }
\newcommand{\W}{ {{\boldmath{W}}} }
\newcommand{\noise}{ {{\boldmath{\eta}}} }
\newcommand{\w}{\omega}
\newcommand{\proposed}{{\tt {wLASSO}}}
\newcommand{\lasso}{{\tt {LASSO}}}

\newcommand{\FC}{{FC}}

\begin{document}

\title{Compressed Wideband Spectrum Sensing: Concept, Challenges and Enablers}

\author{Bechir~Hamdaoui,~\IEEEmembership{Senior~Member,~IEEE,} Bassem~Khalfi,~\IEEEmembership{Student~Member,~IEEE,}
        and~Mohsen~Guizani,~\IEEEmembership{Fellow~Member,~IEEE}}

\maketitle
\begin{abstract}
Spectrum sensing research has mostly been focusing on narrowband access, and not until recently have researchers started looking at wideband spectrum.
Broadly speaking, wideband spectrum sensing approaches can be categorized into two classes:
Nyquist-rate and sub-Nyquist-rate sampling approaches.
Nyquist-rate approaches have major practical issues that question their suitability for realtime applications; this is mainly because their high-rate sampling requirement calls for complex hardware and signal processing algorithms that incur significant delays. Sub-Nyquist-rate approaches, on the other hand, are more appealing due to their less stringent sampling-rate requirement.
Although various concepts have been investigated to ensure sub-Nyquist rates, compressive sampling theory is definitely one concept that has attracted so much interest. This paper explains and illustrates how compressive sampling has been leveraged to improve wideband spectrum sensing by enabling spectrum occupancy recovery with sub-Nyquist sampling rates. The paper also introduces new ideas with great potential for further wideband spectrum sensing enhancements, and identifies key future research challenges and directions that remain to be investigated.
\end{abstract}

\section{Introduction}
\label{sec:introduction}
Spectrum sensing has been the focus of lots of research due to its vital role in promoting dynamic spectrum access. The literature focus has, however, mostly been on narrowband access, and not until recently has wideband spectrum access attracted some momentum, merely due to recent high demands for spectrum resources coupled with the emergence of IoT and 5G technologies, forcing regulatory agencies like FCC to open up new band use in higher frequencies~\cite{FCC-mmW-16}. Although these new regulations and rules bring new opportunities for spectrum access to meet new demands, they also present new spectrum sensing challenges.

Conventional approaches for wideband spectrum sensing consist of first using analog-to-digital converters (ADC) to digitize the wideband signal and then apply digital signal processing (DSP) techniques to locate spectrum vacancy. One simple approach is frequency sweeping, which essentially divides wideband frequency into multiple narrowbands, and then uses narrowband sensing approaches to sweep through all narrowbands to locate spectrum availability. One major issue with this approach is sweeping delay, which can present a great limitation, especially for realtime applications. Another approach is to use multiple filtering hardware blocks, one for each narrowband, to allow parallel sensing across all narrowbands. Though addresses the delay issue, this approach can be very costly from a hardware viewpoint.
Wavelet techniques have also been proposed for performing wideband sensing, which use power spectrum density analysis to detect irregularities that can then be used to locate spectrum availability.
A more natural approach is to sample the time-domain signal occupying the entire wideband at Nyquist (or above) rates and then use FFT methods to determine frequency occupancy across the entire spectrum. Although seems more natural, the issue with these Nyquist-rate sampling approaches is that they require complex hardware and ADC circuitry that have to operate at high sampling rates, as well as sophisticated DSP algorithms that can incur significant delays, making these approaches unpractical when applied to wideband spectrum sensing.

Because of these aforementioned issues, many works have focused on leveraging compressive sampling theory to take advantage of the signal sparsity in the frequency domain to develop wideband spectrum sensing solutions that require sampling rates lower than Nyquist rates~\cite{sharma2016application,sun2013wideband}.
In this paper, we focus on these compressive sampling based spectrum sensing approaches. We first begin by explaining and illustrating how compressive sampling has been leveraged to enable wideband spectrum occupancy recovery at sub-Nyquist sampling rates (Section \ref{sec:compressed}). We then propose new techniques that exploit occupancy heterogeneity in wideband access (Section \ref{sec:weighted}) and cooperative approaches that exploit machine learning (Section \ref{sec:cooperative}) to provide further enhancements to spectrum sensing recovery efficiency. We also identify and present key challenges and future research directions that remain to be investigated (Section~\ref{sec:conc}).
We want to mention that the paper is tutorial in nature and is by no means intended to provide a survey on the topic; it rather starts from key works in the literature that played a vital role in motivating the use of the compressive sampling theory in the context of wideband spectrum sensing~\cite{sharma2016application}, as well as on the authors' own work on the subject to bring the readers' attention to some potentials that remain to be exploited and to some possible ways of exploiting them~\cite{khalfi2017exploiting-twc,khalfi2017machine}.

\section{Compressed Wideband Spectrum Sensing}
\label{sec:compressed}
Consider a wideband system with $\n$ non-overlapping narrowbands, and a secondary user (\SU) receiving primary users' (\PU s) signals that are occupying the entire wideband spectrum. Our goal here is for the \SU~to know/acquire spectrum occupancy of each of the $n$ narrowbands through spectral analysis of its received signal, $\boldsymbol{r}(t)$.

\subsection{Uncompressed spectrum occupancy information recovery}
From Nyquist/Shannon sampling theory, in order to reconstruct $\boldsymbol{r}(t)$ without aliasing, samples with at least twice the maximum wideband frequency, $f_{max}$, must be taken.
Let us consider a sensing window\footnote{Throughout, we assume that the window is chosen small enough that bands' occupancy statuses remain unchanged during such a period.} $[0,\nsamples T_0]$ with $T_0=1/(2f_{max})$, where $\nsamples$ here represents the minimum number of samples needed to guarantee that the signal is sampled at or above the Nyquist rate. The sample vector is the discrete vector $\boldsymbol{r}[l]$ whose $\nsamples$ elements are $\boldsymbol{r}[l]=\boldsymbol{r}(t)|_{t=lT_0}, l=0,1,\ldots,\nsamples-1$. One obvious spectrum occupancy recovery approach would consist of performing a discrete Fourier transform (DFT) on the sample vector to compute the energy level present in each of the narrowbands, and then use these computed energy values to decide on narrowbands' availabilities.
More specifically, the received signal occupying narrowband $b$, $b=1,2,\ldots,\nsamples$, can be represented in the frequency domain by its DFT to be calculated using $\boldsymbol{r}[l]$; i.e.,  $R_b=\frac{1}{\nsamples} \sum_{l=0}^{\nsamples-1}{\boldsymbol{r}[l] e^{-j2\pi b l/\nsamples}}$.
Now for each narrowband $b$, one can repeat this process $M$ times over different intervals, compute the sum statistics of the received energy on that narrowband (i.e., $\sum_{t=1}^{M}{|R_b[t]|^2}$), and compare it against some threshold to decide whether narrowband $b$ is available. Note that the larger the $\nsamples$, the longer the sensing period and hence the greater the number of taken samples, but also the greater the number of sampled frequencies (i.e., the better the resolution). Throughout, for normality and simplicity, we consider $\nsamples=n$; i.e., the number of sampled frequencies is set to the number of narrowbands.

As mentioned earlier, the challenge with this uncompressed signal recovery approach is that it requires high sampling rates, thus calling for complex ADC hardware and signal processing algorithms. This prompted researchers to look for compressed approaches as alternative solutions.

\subsection{Compressed spectrum occupancy information recovery}
Various measurement studies reveal that the wideband spectrum has relatively low occupancy~\cite{yilmaz2016determination}, thereby allowing to leverage compressive sampling to recover spectrum occupancy information with sub-Nyquist sampling rates~\cite{tian2007compressed}. Briefly said, compressive sampling theory allows to reconstruct signals (or vectors) that are {\em sparse} through sampling rates that are (much) lower than Nyquist rates, where, formally, a vector $\boldsymbol{x}\in\mathbb{R}^n$ is said to be \emph{k-sparse} if it has (with or without a basis change) at most $k$ non-zero elements; i.e., $supp(\boldsymbol{x}):=\|\boldsymbol{x}\|_{\ell_0} = |\{i : \boldsymbol{x}[i]\neq 0\}|\leq k$.
In our wideband sensing application case, letting $\boldsymbol{x}$ be the ${n\times 1}$ vector representing the occupancy information of the $n$ narrowbands
(with $0$ being vacant),
the sparsity $k$ of $\boldsymbol{x}$ refers to the number of occupied narrowbands.
Because of this sparsity, compressive sampling comes then handy and allows to recover occupancy information captured via the length-$n$ vector $\boldsymbol{x}$ with only $m \ll n$ measurements~\cite{davenport2011introduction}. Throughout, $\boldsymbol{y}$ will denote the length-$m$ vector of these $m$ measurements.

\subsubsection{Compressed spectrum sensing}
Recall that the discrete vector $\boldsymbol{r}$ whose elements are the samples of the received signal
$\boldsymbol{r}(t)$ at $t=lT_0$, $l=0,1,\ldots,\n-1$, can be expressed in terms of the $n \times 1$ inverse-Fourier basis column vectors $\{{\dfti_i}\}_{i=0}^{n-1}$ as
$\boldsymbol{r}=\sum_{i=0}^{n-1}{\dfti_i \boldsymbol{x[i]}}$ or in matrix notation as $\boldsymbol{r}=\dfti \boldsymbol{x}$, where $\boldsymbol{x}$ is again the ${n\times 1}$ vector representing the occupancy information of the $n$ narrowbands during the corresponding sensing period, and $\dfti$ is the $n \times n$ matrix whose columns are $\{{\dfti_i}\}_{i=0}^{n-1}$. Note that although either $\boldsymbol{x}$ or $\boldsymbol{r}$ suffices for uniquely representing the Nyquist-rate samples of the received signal, only the frequency-domain representation $\boldsymbol{x}$ is sparse. That is, using $k$ to again refer to the sparsity level, $\boldsymbol{r}$ can then be viewed as a linear combination of only $k\ll n$ columns of the basis matrix $\dfti$. It is this sparsity structure that allows for the use of compressive sampling to recover $\boldsymbol{x}$ with only $m\ll n$ samples as opposed to all $n$ samples. Letting the $m\times n$ matrix $\measure$ represent the $n$-to-$m$ reduction matrix with $m$ length-$n$ rows $\{\measuree_j\}_{j=0}^{m-1}$, one can write the length-$m$ vector $\boldsymbol{y}$ of these $m$ measurements as $\boldsymbol{y}=\measure \boldsymbol{r}$, or alternatively,
$\boldsymbol{y}=\sense \boldsymbol{x}$
with $\sense = \measure \dfti$ by replacing $\boldsymbol{r}$ by $\dfti \boldsymbol{x}$. Here each measurement $y_i=\langle\measuree_i,\boldsymbol{r}\rangle$ is nothing but a linear combination of the $n$ samples.
First, note that recovering $\boldsymbol{x}$ by solving the system $\boldsymbol{y}=\sense \boldsymbol{x}$ given $\boldsymbol{y}$ would be an ill-posed problem had $\boldsymbol{x}$ not been sparse, since there would be more unknowns than equations. Because $\boldsymbol{x}$ is $k$-sparse, it is then possible to recover it from only $m$ measurements (i.e., $\boldsymbol{y}$) provided that $\sense$ possesses the {\em restricted isometry property} (RIP)~\cite{candes2006robust}, which essentially means that every set of $k$ of fewer columns of $\sense$ behaves approximately like an orthonormal system.
Therefore, one fundamental question that has attracted significant research attention is how to construct the reduction matrix $\measure$ such that $\sense$ possesses the RIP? It has been shown that a Gaussian matrix $\measure$ whose elements $\measure_{i,j}$ are each drawn from an i.i.d. Gaussian random variable with zero mean and $1/m$ variance ensures, with an overwhelming probability, that the matrix $\sense=\measure \dfti$ has the RIP and that the vector $\boldsymbol{x}$ can be recovered with just $m=O \big(k\log(n/k)\big)$ measurements~\cite{donoho2006compressed}.

\subsubsection{Spectrum recovery approaches}
When considering a noise-free environment, one can recover $\boldsymbol{x}$ by simply finding $\boldsymbol{z}$ that minimizes
$\|\boldsymbol{z}\|_{\ell_0}$ subject to
$\boldsymbol{y}=\Psi\boldsymbol{z}$. However, solving such a combinatorial problem is computationally expensive, and as a result, heuristic approaches (e.g., BP~\cite{candes2006robust} and OMP~\cite{tropp2007signal}) have been proposed as an alternative solution. For instance, it has been shown that this combinatorial problem can equivalently be formulated as a convex optimization problem (and hence can be solved via classical linear programming) by simply minimizing the $\ell_1$-norm of $\boldsymbol{z}$ instead of its $\ell_0$-norm; this is widely known in the literature as the Basic Pursuit (BP)~\cite{candes2006robust}.

In practice, the $m$ measurements (i.e., $\boldsymbol{y}$) from which we intend to recover our spectrum occupancy information vector $\boldsymbol{x}$ are often not noise free. Let us be more specific by considering a faded and noisy communication environment, in which, the discrete signal $\boldsymbol{r}[l]$, $l=0,1,\ldots,\n-1$, sampled at the \SU's front-end, can be expressed as $\boldsymbol{r}[l]=\boldsymbol{h}[l]* \boldsymbol{s}[l]+\boldsymbol{w}[l]$, where $\boldsymbol{h}[l]$ is the channel impulse between primary transmitters and the \SU, $\boldsymbol{s}[l]$ is the \PU~transmitted signal, $\boldsymbol{w}[l]$ is an Additive White Gaussian Noise with $\boldsymbol{w}[l] \sim N(0,\sigma^2)$, and $*$ is the convolution operator. Now performing a discrete Fourier Transform on the expression of the received discrete signal $\boldsymbol{r}$ yields ${R}={H}{S}+{W}=\boldsymbol{x}+{W}$,
where ${H}$, ${S}$, and ${W}$ are the Fourier transforms of $\boldsymbol{h}[l]$, $\boldsymbol{s}[l]$, and $\boldsymbol{w}[l]$, respectively, and then performing the inverse Fourier transform on the obtained equation yields
$\boldsymbol{r}=\dfti R=\dfti \boldsymbol{x} + \dfti W$.
The vector $\boldsymbol{x}$ here contains faded versions of \PU s' signals sent at the different narrowbands.
Now given that the measurement vector $\boldsymbol{y}=\measure \boldsymbol{r}$, we can then write $\boldsymbol{y}=\measure \dfti \boldsymbol{x} + \measure \dfti W$, or more compactly,
$\boldsymbol{y}=\sense \boldsymbol{x} + \boldsymbol{\noise}$
with $\boldsymbol{\noise} = \sense W$, where again $\sense = \measure \dfti$ and  $\measure$ is the $n$-to-$m$ reduction matrix that reduces the number of measurements/taken samples from $n$ down to only $m$. Unlike in the case of the noise-free (ideal) environment, in this noisy (realistic) environment, not only do we have fewer samples of the signal, but also these few observations are not accurate. Fortunately, compressive sampling theory comes handy and can help recover $\boldsymbol{x}$ even in this imperfect setting. Clearly, the recovered vector cannot be exact now due to the imperfection of the collected measurements. However, many recovery approaches with various bounds on the error have been developed for this specific scenario. The $\ell_1$-minimization approach~\cite{candes2006stable}, commonly known as LASSO, and greedy pursuits like OMP~\cite{tropp2007signal}, CoSaMP~\cite{needell2009cosamp} and AS-SaMP~\cite{zhang2016novel} are good representatives of such approaches.
For instance, LASSO~\cite{candes2006stable} finds, among all feasible signals, the sparsest one with a bounded error by solving the following $\ell_1$-minimization problem ($\mathscr{P}_{\lasso}$):
\begin{equation}\label{eq:lasso}
\mathscr{P}_{\lasso}:\hspace{0.2in} \underset{\boldsymbol{z}}{\text{minimize}} \|\boldsymbol{z}\|_{\ell_1} \hspace{0.2in} \text{subject to} \hspace{0.2in} \|\sense \boldsymbol{z}-\boldsymbol{y}\|_{\ell_2}\leq \epsilon
\end{equation}
where $\epsilon \geq \|\boldsymbol{\eta}\|_{\ell_2}$ is a pre-defined parameter.

\subsubsection{Hardware implementation} \label{subsec:hw}
\begin{figure}
\centering{
\includegraphics[width=0.9\columnwidth]{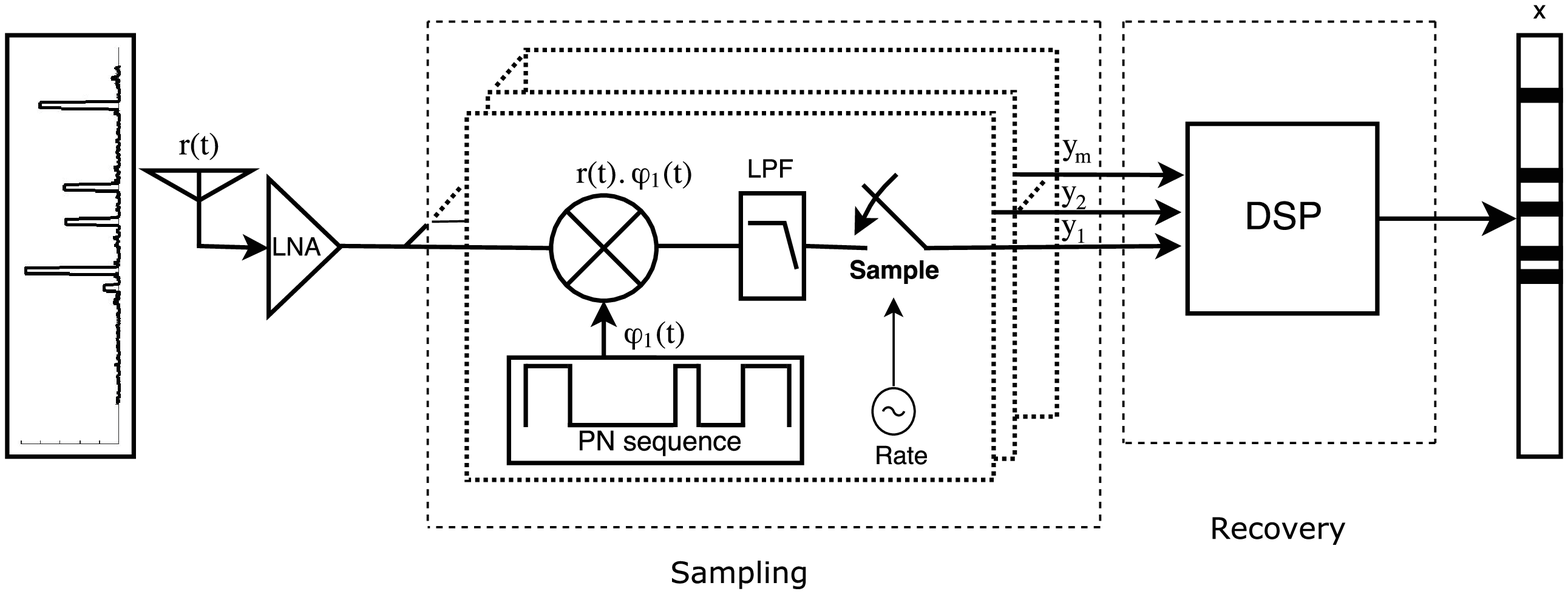}
\caption{Compressed wideband spectrum sensing architecture.}
\label{fig:implement}}
\end{figure}

Recently, there have also been some efforts aimed at designing new hardware architectures suitable for compressed wideband spectrum sensing (e.g.,~\cite{yazicigil2015wideband}), with an overall focus on balancing among scanning time, energy consumption and hardware complexity/cost.
For illustration, Fig.~\ref{fig:implement} shows a high-level implementation capturing the key components of these architectures.
First, the received wideband RF signal $\boldsymbol{r}(t)$ is amplified using a low noise amplifier (LNA) and fed to $m$ parallel branches, where at each branch, $\boldsymbol{r}(t)$ is mixed with a unique pseudo-random (PN) sequence (e.g., of $\pm1$).
The mixing step at each branch essentially modulates the signal $\boldsymbol{r}(t)$ with a $n$-length random signal $\varphi_i(t)$, resulting in a signal $r(t)\varphi_i(t)$ that is nothing but a linear combination of shifted copies of frequency-domain signals occupying each band of the wideband spectrum. In other words, the mixing operation spreads the entire spectrum so that the low-pass filtered (LPF) output of each branch is a narrow band copy of the signal that contains energy from all the other bands.
Connecting this with the theory we discussed previously, each PN sequence can be viewed as one row of the sensing matrix, and hence, it is important that the PN sequences are uncorrelated to ensure reliable recovery.
After low-pass filtering, sampling is then performed at each branch $i$ at a rate (much lower than Nyquist rate) determined by the width of the narrow band, resulting in an output sequence, ${y}_i[n]$. Again, here the frequency-domain version of each sequence $y_i[n]$ is a combination of the shifted versions of signals occupying the different bands.
Finally, a DSP algorithm is used to recover the signal and provide the occupancy of every band.

\section{Weighted Compressed Wideband Spectrum Sensing}
\label{sec:weighted}

\begin{figure}[t]
\centering{
\includegraphics[width=.9\columnwidth]{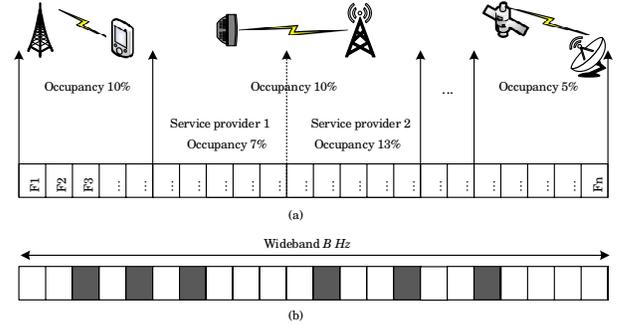}
\caption{Spectrum bands occupied by applications with different occupancies. Grey bands are occupied by primary users and white bands are vacant. (a) average occupancies of the different spectrum blocks. (b) one possible allocation at some time slot.}
\label{fig:band_ocup}}
\end{figure}

In practice, applications of similar types (cellular, satellite, TV, etc.) are often assigned spectrum bands within the same (or nearby) frequency block.
Also, different application types may show different occupancy patterns and characteristics. These two facts lead to an important observation (also supported via measurements~\cite{yilmaz2016determination}): different frequency blocks exhibit different occupancy statistics (see Fig.~\ref{fig:band_ocup}). Throughout, we refer to this variability in spectrum bands' occupancies across the different blocks as {\em block-like spectrum occupancy structure}.

{\bf {\em The proposed \proposed:}}
In this section, we present an approach~\cite{khalfi2017exploiting-twc} that exploits this
block-like structure to improve the recovery efficiency, in terms of error bounds and number of required measurements, of the spectrum occupancy information vector $\boldsymbol{x}$. To ease the illustrations, we assume that the $n$ narrowbands are grouped into $g$ disjoint contiguous blocks, $\mathcal{G}_i, i=1,...,g$, with $\mathcal{G}_i\bigcap\mathcal{G}_j=\emptyset$ for $i\neq j$, with each block, $\mathcal{G}_i$, consisting of $n_i$ contiguous bands.
For simplicity, we model the state of each band $i$, $\mathcal{H}_i$, as $\mathcal{H}_i \thicksim \text{Bernoulli}(p_i)$
with parameter $p_i\in[0,1]$ where $p_i$ is the probability that band $i$ is occupied by some \PU.
Let $\bar{k}_j= \sum_{i\in \mathcal{G}_j } p_i$ be the average number of bands occupied within block $j$ (assuming independency across band occupancies). The block-like structure of spectrum occupancy behavior dictates that $\bar{k}_j$ varies from one block to another; when necessary, blocks with similar sparsity levels are merged together and assigned a sparsity level that corresponds to their average. These per-block spectrum occupancy averages can be directly estimated via measurements or provided by spectrum operators~\cite{yilmaz2016determination}.
Our proposed recovery approach, referred to as {\em weighted LASSO (\proposed)}, incorporates and exploits the sparsity variability observed across the different frequency blocks to allow for a more efficient solution search.
Referring to $\mathscr{P}_{\lasso}$ (Equation~\eqref{eq:lasso}) again for illustration, let's write the vector variable $\boldsymbol{z}$ as $\boldsymbol{z}=[\boldsymbol{z}_1^T,\boldsymbol{z}_2^T,\ldots, \boldsymbol{z}_g^T]^T$ where $\boldsymbol{z}_i$ is the $n_i\times 1$ vector corresponding to block $i$ for $i\in\{1,...,g\}$, and assign for each block $i$ a weight $\w_i$ such that $\w_i>\w_j$ when $\bar{k}_i < \bar{k}_j$ for all blocks $i,j$. Essentially, the weights are designed in such a way that
a block with higher sparsity level is assigned a smaller weight; for instance, setting $\omega_i=(1/\bar{k}_i)/{\sum_{j=1}^{g}(1/\bar{k}_j)}$ meets such a design requirement. The proposed \proposed~is then:
\begin{equation}\label{eq:wlasso}
\mathscr{P}_{\proposed}: \hspace{0.05in} \underset{\boldsymbol{z}}{\text{minimize}} \sum_{i=1}^g \omega_i\|\boldsymbol{z}_{i}\|_{\ell_1} \hspace{0.05in} \text{subject to} \hspace{0.05in} \|\sense\boldsymbol{z}-\boldsymbol{y}\|_{\ell_2}\leq \epsilon
\end{equation}
Intuitively, by assigning smaller weights to blocks with higher sparsity levels, \proposed~ensures that the search for a sparse solution vector, among all feasible vectors, is aimed towards lesser sparse blocks, thereby $(i)$ reducing recovery errors and/or $(ii)$ requiring lesser numbers of measurements~\cite{khalfi2017exploiting-twc}.

\begin{figure}
\centering{
\includegraphics[width=0.9\columnwidth]{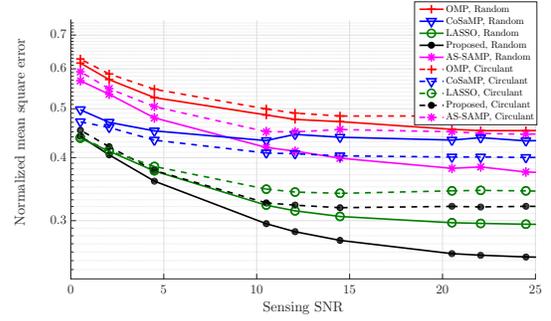}
\caption{Normalized Mean Square Error ($\|\boldsymbol{z}^*-\boldsymbol{x}\|_{\ell_2}/\|\boldsymbol{x}\|_{\ell_0}$) as a function of received signal SNR ($\|\boldsymbol{x}\|_{\ell_2}^2/\|\boldsymbol{\eta}\|_{\ell_2}^2$) for $m=27$.}
\label{fig:perfSNR}}
\end{figure}

\begin{figure}
\centering{
\includegraphics[width=0.9\columnwidth]{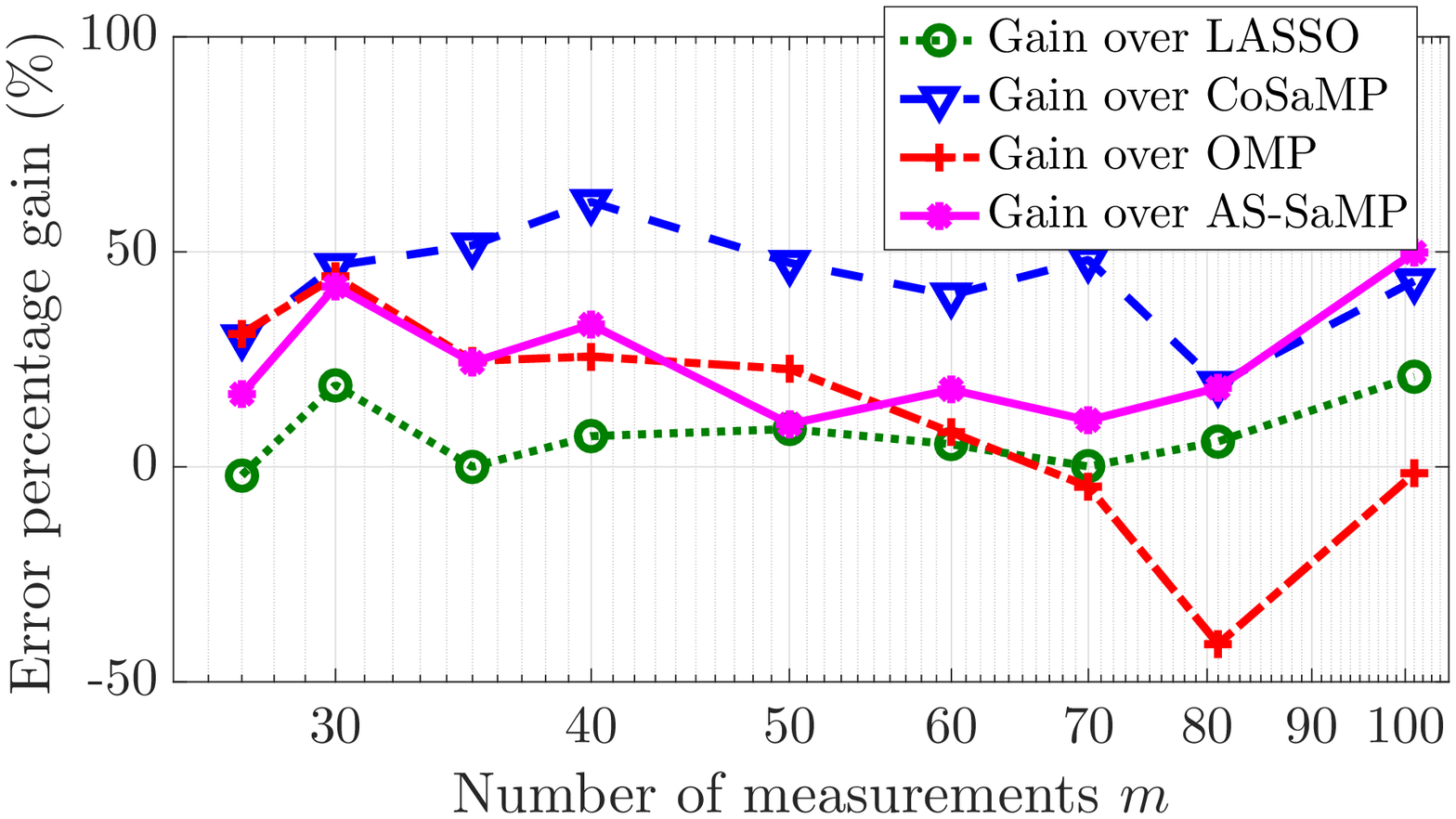}
\caption{Error gain (defined as ($Error_{other}-Error_{\proposed})/Error_{other}$) as a function of $m$ for SNR$=7dB$ and for random sensing matrix.}
\label{fig:perfm}}
\end{figure}

{\bf {\em Performance analysis of \proposed:}}
Figs.~\ref{fig:perfSNR} and~\ref{fig:perfm} show error performances achieved under proposed \proposed, LASSO~\cite{candes2006stable}, OMP~\cite{tropp2007signal}, CoSaMP~\cite{needell2009cosamp} and AS-SaMP~\cite{zhang2016novel} for random Bernoulli and Circulant sensing matrices, and Table~\ref{tab1} provides their complexity analysis. Three observations can be made from these results: One, \proposed~incurs the smallest errors because it encourages the search to take place in the portions of the spectrum with more occupied bands.
Therefore, with the same number of measurements, \proposed~yields accuracy higher than LASSO, and does so without compromising its computational complexity. This is because the difference between them is that \proposed~assigns different weights to different blocks, whereas LASSO assigns equal weights to all blocks, and hence, solving these two algorithms take the same amount of time. However, this gain comes at the cost of needing to know the average of occupancy of each block in advance. But when compared to OMP, CoSaMP and AS-SaMP, \proposed's error gain comes also at a higher computational complexity.
Two, the random sensing matrix always incurs lesser errors regardless of the recovery approach being used. In essence, to achieve a robust recovery, the rows of the sensing matrix should have low cross-correlation, which is achieved more with a fully random matrix.
Three, the error gains of \proposed~are maintained over LASSO, CoSaMP and AS-SaMP but not over OMP when the number of measurements is high. However, as can be recalled from Fig.~\ref{fig:implement}, having a large number of measurements is not of interest since it requires more branches, and hence, a more costly hardware.

\begin{table*}
  \centering
    \caption{Performance comparison among recovery approaches}  \label{tab1}
\begin{tabular}{|l|l|l|l|}
  \hline
  \bf{Approach} & \bf{Complexity}  & \bf{Limitations} & \bf{Strengths} \\\hline \hline
  OMP~\cite{tropp2007signal} & $O(mnk)$ & No guarantees for noisy and compressible signals & Fast\\ \hline
  CoSaMP~\cite{needell2009cosamp} & $O(mn\;itr)$ &No guarantees for noisy and compressible signals  & Fast (slower than OMP) but better performance \\ \hline
  AS-SaMP~\cite{zhang2016novel} & $O(mn\;itr)$ & No guarantees for noisy and compressible signals  & Faster than CoSaMP but slower than OMP \\ \hline
  LASSO~\cite{candes2006stable} & $O(m^2n^3)$ & Slow convergence for high $n$ and $m$ & Provable guarantees for stable recovery \\ \hline
  \proposed~\cite{khalfi2017exploiting-twc} & $O(m^2n^3)$ & Slow convergence for high $n$ and $m$,& Provable guarantees for stable recovery  \\
  &  & requires some a priori knowledge for weights' design& \\
  \hline
\end{tabular}
\end{table*}

{\bf {\em Hardware implementation of \proposed:}}
As described in Section~\ref{subsec:hw}, \proposed~can also be implemented by first mixing the amplified wideband RF signal $\boldsymbol{r}(t)$ with $m$ different PN sequences of $\pm 1$, sampling each of the $m$ low-pass filtered signals, and then digitally solving the weighted optimization at the DSP level to recover the occupancy information.
Another approach of implementing \proposed~is to implement the weighted compression at the RF front-end instead of being done at the DSP end. For this, observe that $\proposed$, formulated in Eq.~\eqref{eq:wlasso}, could equivalently also be reformulated as to minimize $\|\boldsymbol{z}\|_{\ell_1}$ subject to $\|\sense \Winv \boldsymbol{z}-\boldsymbol{y}\|_{\ell_2}\leq \epsilon$
where $\W=\textrm{diag}(\underbrace{\w_1,\cdots,\w_1}_{n_1},\underbrace{\w_2,\cdots,\w_2}_{n_2},\dots,\underbrace{\w_g,\cdots,\w_g}_{n_g})$. %
The new sensing matrix $\Winv$ essentially magnifies the columns of the sensing matrix $\sense$ that correspond to high average sparsity levels (low weights) and belittles the columns that correspond to low average sparsity levels.

\section{Cooperative and Adaptive Compressed Wideband Spectrum Sensing}
\label{sec:cooperative}
In this section, we focus on three practical issues.
One, bands' occupancies are time varying. That is, not only does spectrum occupancy vary from one frequency block to another, but also over time. In other words, the block sparsity levels $\bar{k}_j$s are not fixed. It is therefore important to devise adaptive approaches that can provide accurate estimates of these levels.
Being able to have accurate estimates of sparsity levels is vital so that the number of measurements needed for the recovery can be determined accurately; remember over-sampling (using more measurements than needed) incurs greater overheads, whereas under-sampling leads to inaccurate recovery.
Two, recall that the number, $m$, of hardware branches needed by a \SU~device to perform \proposed~depends on the sparsity level, $\bar{k}_j$, of each block $j$. Since $\bar{k}_j$ varies over time, then so does $m$. On the other hand, the number of hardware branches a receiver can have can only be fixed and is often way smaller than $m$. Therefore, there is a need for adaptive approaches that address the limited number of hardware branches as well as the variability of $m$.
Three, because an \SU's ability to detect a \PU's signal depends on its distance from the \PU~(among other things), an \SU's signal recovery may be erroneous. This problem---aka the hidden terminal problem---needs also to be carefully addressed.

In this section, we present two approaches that address the three aforementioned issues. Specifically, we rely on prediction to overcome the first issue, and on cooperation to overcome the second and third issues.

\subsection{Spectrum occupancy prediction} \label{subsec:prediction}
\begin{figure}
\centering{
\includegraphics[width=0.9\columnwidth]{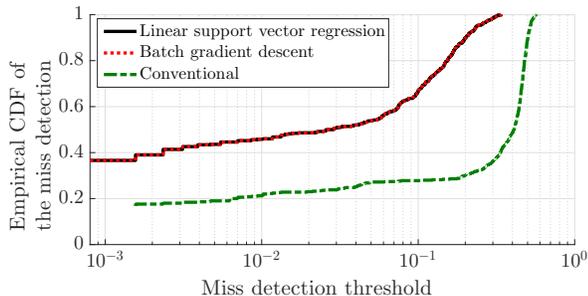}
\caption{Empirical Cumulative Distribution Function (CDF) of miss detection: prediction-based vs. conventional (no prediction).}
\label{fig:SE}}
\end{figure}

One way to overcome the sparsity level time-variability issue is to incorporate and rely on prediction models to track and provide accurate estimates of these occupancy levels.
Fig.~\ref{fig:SE} illustrates how the performance of the weighted compressed sensing approach behaves with and without prediction when considering two regression models: batch gradient descent and linear support vector regression~\cite{khalfi2017machine}.
The figure shows that prediction leads to a more accurate recovery (low miss detection rate of occupied spectrum bands), as these regression schemes allow to provide more accurate numbers of the required measurements.
Due to space limitation, here we skipped the simulation/evaluation setup details; these details, as well as more result insights and  prediction analysis, can be found in~\cite{khalfi2017machine}.

\subsection{Cooperative wideband spectrum sensing} \label{subsec:cooperative}
User cooperation can be leveraged to address the hidden terminal problem, as well as the time-variability of $m$ and the limitation in the number of hardware branches.
To address the hidden terminal problem, one can first have each \SU~take and report $m$ local measurements to a fusion center (\FC). Then, \FC, after recovering the occupancy information by applying \proposed~on each of these \SU s' local measurements, uses a voting mechanism to decide on whether the bands are occupied. Alternatively, each \SU~can apply \proposed~locally and send the occupancy vector instead of sending the measurements. In this case, \FC~can run the voting and decide on the spectrum band availability without needing to apply \proposed~on each measurement set. This cooperative sensing can be (and is often) used as a way for addressing the hidden terminal problem.
More details on this can be found in~\cite{khalfi2017machine}.

Now for addressing the time-variability of $m$ and the constrained number of hardware branches, our approach first uses prediction techniques (e.g, those described in Section~\ref{subsec:prediction}) to estimate $m$, and then have each \SU~perform one sensing scan using whatever (limited) number of branches it has and send its measurement vector to \FC. Note that an \SU~can choose to perform multiple (sequential) sensing scans via its hardware, leading to more measurements (but also to more delay). Hence, the measurement vector size depends on the \SU's number of branches and number of performed scans.
When the number of combined measurements received by \FC~reaches $m$, \FC~applies \proposed~to recover the spectrum occupancy information.

\section{Open Research Challenges}\label{sec:conc}
Although, as explained in this paper, compressive sampling shows great potential for improving wideband spectrum sensing, there still remains key challenges that, when addressed, further enhancements can be achieved:
\begin{itemize}
  \item {\bf Signal type identification.}
  Most spectrum sensing literature focused on detecting whether bands are occupied or not, but not so much on identifying what types of signals/transmitters are occupying them. Signal identification research has mainly focused on RF fingerprinting, which basically tries to extract features (e.g., modulation type) that are intrinsic to the transmitted signals and/or device manufacturing imperfections, and use these features to discriminate among the transmitters. The problem is that most of these fingerprinting techniques can only discriminate among devices that are identical (maybe produced by different manufacturers) and that operate using one type of communication protocol (e.g. WiFi). With the rapid emergence of IoT, multiple types of wireless protocols will emerge and possibly coexist, thereby calling for more sophisticated discriminatory approaches. Being able to identify and discriminate among different types of devices operating using different protocols will be vital to spectrum access enforcement. We expect that machine learning will play a key role in helping develop automated algorithms that allow to identify key features and classify signals.
  \item {\bf Wideband spectrum databases.}
  As done for the TV bands, there has recently also been a consensus for the need of databases that serve widebands. However, unlike the case of TV white space databases, building such databases presents new requirements and challenges, specific to the next-generation spectrum access systems at hand. For instance, the massive numbers of IoT devices that are expected to coexist and need spectrum resources, though give rise to obvious resource bottleneck challenges, can surely be leveraged to improve spectrum sensing reliability and overhead. This, for example, can be done through the use of collaborative filtering, a theory that has already been successfully adopted in domains like recommendation systems, and can surely be exploited to build such databases.

  \item {\bf Adaptive hardware.}
  As discussed in Section~\ref{subsec:hw}, there have already been proposed new hardware architectures suitable for compressed wideband sensing, with a focus on reducing sensing time and energy consumption while keeping hardware cost at minimum. One key challenge with these existing architectures is that they do not adapt to signals' sparsity levels.
  This is because the number of hardware branches can only be fixed and is often way smaller than $m$.
  Besides, this number $m$ changes over time, as it depends on the time-varying spectrum occupancy.
  Therefore, although we discussed in Section~\ref{subsec:cooperative} some high-level cooperative approaches for dealing with such a problem, there remains a need for new solutions at the hardware level that can adapt to sparsity levels in real time so that reliable recovery is guaranteed when actual sparsity levels vary.
\end{itemize}

\section{Acknowledgement}
This work was supported in part by the US National Science Foundation (NSF) under NSF award CNS-1162296.

\bibliographystyle{IEEEtran}
\bibliography{References-new,References-twc}
\begin{IEEEbiographynophoto}
{Bechir Hamdaoui (S'02-M'05-SM'12)}
is an Associate Professor in the School of EECS at Oregon State University. He received his Ph.D. in ECE from University of Wisconsin at Madison (2005). His research interests span various topics in the fields of computer networking and wireless communication. He serves/served on the editorial boards of few journals and as the program chair/co-chair and TPC of many conferences. Dr. Hamdaoui is an IEEE Senior Member.
\end{IEEEbiographynophoto}
\begin{IEEEbiographynophoto}{Bassem Khalfi (S'14)} %
is currently a Ph.D. candidate in ECE at Oregon State University. He received his Graduate Engineer Diploma (2012) from the Higher School of Communication of Tunis (SUP'COM) in Ariana, Tunisia, and his M.S. degree (2014) from the National School of Engineers, Tunis, Tunisia.
His research focuses on various topics in the area of wireless communication and networks, including dynamic spectrum access and sensing, RF energy harvesting, and IoT.
\end{IEEEbiographynophoto}
\begin{IEEEbiographynophoto}
{Mohsen Guizani (S'85-M'89-SM'99-F'09)}
 is currently a Professor and the Chair of the ECE Department at the University of Idaho, USA. He received his Ph.D. in Computer Engineering in1990, from Syracuse University. His research interests include computer networks and wireless communications. He currently serves on the editorial boards of six Journals. Dr. Guizani is an IEEE Fellow and a Senior member of ACM.
\end{IEEEbiographynophoto}

\end{document}